\documentclass[twocolumn,secnumarabic,amssymb, nobibnotes,aps,prl]{revtex4}

\newcommand{\env}[1]{\texttt{#1}}
\usepackage{graphicx,url,epstopdf,color}
\usepackage{amsmath,amsthm,amssymb}

\renewcommand{\vec}[1]{\boldsymbol{#1}}

\definecolor{Black}{named}{Black}
\definecolor{Blue}{named}{Blue}
\definecolor{Red}{named}{Red}

\def\fun#1#2{\lower3.6pt\vbox{\baselineskip0pt\lineskip.9pt
  \ialign{$\mathsurround=0pt#1\hfil##\hfil$\crcr#2\crcr\sim\crcr}}}

\newcommand{\be}{\begin{equation}}
\newcommand{\ee}{\end{equation}}
\newcommand{\ba}{\begin{eqnarray}}
\newcommand{\ea}{\end{eqnarray}}

\def\ap{\approx}

\def\lsim{\raise0.3ex\hbox{$\;<$\kern-0.75em\raise-1.1ex\hbox{$\sim\;$}}}
\def\gsim{\raise0.3ex\hbox{$\;>$\kern-0.75em\raise-1.1ex\hbox{$\sim\;$}}}

\def\theta{\vartheta}

\begin{document}

\title{Filamentary Diffusion of Cosmic Rays on Small Scales}

\author{G.~Giacinti$^{1}$}
\author{M.~Kachelrie\ss$^{1}$}
\author{D.~V.~Semikoz$^{2,3}$}
\affiliation{$^1$Institutt for fysikk, NTNU, Trondheim, Norway}
\affiliation{$^2$AstroParticle and Cosmology (APC), Paris, France}
\affiliation{$^{3}$Institute for Nuclear Research of the Russian Academy of Sciences, Moscow, Russia}

\begin{abstract}
We investigate the diffusion of cosmic rays (CR) close to their sources. 
Propagating individual CRs in purely isotropic turbulent magnetic fields 
with maximal scale of spatial variations $l_{\max}$, we find that CRs 
diffuse anisotropically at distances $r\lsim l_{\max}$ from their sources.
As a result, the CR densities around the sources are strongly irregular and show 
filamentary structures. We determine the transition time $t_\ast$
to standard diffusion as 
$t_\ast\sim 10^4 {\rm yr}\; (l_{\max}/150\,{\rm pc})^{\beta} 
(E/{\rm PeV})^{-\gamma}  (B_{\rm rms}/4\,\mu G)^{\gamma}$, with 
$\beta \simeq 2$ and $\gamma=0.25$--$0.5$ for a turbulent field with  Kolmogorov power spectrum. We calculate the photon emission 
due to CR interactions with gas and the resulting irregular source images. 
\end{abstract}


\maketitle




\textit{Introduction.}---The suggestion that Galactic cosmic rays (CR) are accelerated using
the energy released in supernova (SN) explosions dates back to the
1930's~\cite{BZ}. This idea was supported initially mainly by the argument 
that SNe inject sufficient energy into the Galaxy to maintain the 
observed CR energy density, while later the radio emission observed
from  SN remnants (SNR) was interpreted as indication for the acceleration
of high-energy electrons. Until present, a clear proof for both the 
acceleration of hadrons and the identity of their sources
is still missing~\cite{review}. Alternatively, the required power could be 
provided
by acceleration processes which operate at distances
and time scales larger than for individual SNRs, e.g.\ in 
superbubbles~\cite{superbubbles}.

Main obstacle for the identification of CR sources is the diffusion
of CRs in the Galactic magnetic field (GMF), erasing directional 
information on the position of their sources. 
The GMF has a turbulent component which varies on scales between $l_{\min}\lesssim 1$\,AU 
and $l_{\max}\sim{\rm few~to~} 200$\,pc.  Since CRs scatter on inhomogeneities with variation 
scales comparable to their Larmor radius, the propagation of Galactic CRs in the GMF resembles a random walk 
and is well described by the diffusion approximation~\cite{JG99,galprop}.

However, CRs around young sources do not have time to diffuse far away. They 
should produce an extended gamma-ray halo, which can be detected using 
Cherenkov telescopes and gamma-ray satellites. If SNe power the Galactic CR
population, about ten sources with degree extension should be detected at energies 
$E_{\gamma}>100$\,GeV in the Galactic plane assuming the sensitivity of Fermi-LAT, 
while Ref.~\cite{NS12} found 18. Most of these sources were observed also 
as extended sources up to energies $E_{\gamma}\gsim10$\,TeV by the HESS 
experiment~\cite{HESS} and have a non-spherical shape.
Similarly, the Veritas observations
of Tycho show a clear asymmetric extension of TeV photons towards
the north of the SNR~\cite{VERITAS}.

The diffusion approximation cannot predict local phenomena which arise below the CR mean free path $\lambda \ll l_{\max}$, when the local configuration of the turbulent field must lead to observable imprints~\cite{Giacinti:2011mz}. Before the present study it has remained unclear to which extent the diffusion approximation is satisfied on intermediate scales $\lambda\ll l \lesssim l_{\max}$, where large-scale fluctuations of the field lead to local anisotropies, which in turn can explain the irregular images of extended sources found in Refs.~\cite{NS12,HESS,VERITAS}.

We study therefore in this letter the diffusion of CRs on scales comparable to 
the coherence length of the turbulent GMF, ${\cal O}(100\,{\rm pc})$. In 
contrast to earlier studies, we calculate the diffusion tensor propagating 
individual CRs in specific realizations of the turbulent magnetic field. 
We find that diffusion is 
anisotropic even for an isotropic random field and can lead to a filamentary 
structure of the CR density around young sources. 
Responsible for these anisotropies are turbulent field modes with variation 
scales much larger than the Larmor radius of CRs which mimic a regular
field. This effect can be confirmed via the observation of irregular gamma-ray emissions around CR sources.


\textit{Cosmic rays in magnetic turbulence.}---Since we want to test effects beyond the diffusion approximation, we propagate individual CRs in turbulent magnetic fields ${\vec B}({\vec k})\propto \exp(-i\vec k\vec x)$ using the numerical code described in~\cite{nuc2,nuc3}. The validity of this code was checked reproducing earlier results 
from~\cite{Casse:2001be,DeMarco:2007eh}. We assume that the
spectrum $\mathcal{P}(k)$ of magnetic field fluctuations is static and
follows a power-law, $\mathcal{P}(k)\propto k^{-\alpha}$. The former
assumption is justified, because the changes introduced by a finite 
Alfv\'en velocity are small for the considered time scales. 
We fix the mean magnetic field strength
$B_{\rm rms}^2\equiv \langle {\vec B}^2({\vec x})\rangle$ as 
$B_{\rm rms}=4\,\mu$G, normalising $B_{\rm rms}$ for fluctuations bounded by $l_{\min}=1$\,AU and $l_{\max}=150$\,pc. In the numerical simulations, we choose $l_{\min}$ sufficiently small compared to the CR Larmor radius $R_L$.
We also adopt an isotropic spectrum of fluctuations, as we want to demonstrate that 
even in this case CRs diffuse initially anisotropically.

The spectral index of the turbulent GMF is only weakly 
constrained, and both Kolmogorov ($\alpha=5/3$) and Kraichnan 
($\alpha = 3/2$) spectra are consistent with observations~\cite{galprop,obs}. 
As our results do not vary much between these two cases, we present them for a Kolmogorov spectrum. The turbulence is expected to have a Bohm spectrum ($\alpha=1$) only close to shocks, where efficient CR acceleration requires a diffusion coefficient $D(E)\sim R_L$.
As we are interested in time-scales when CRs have already escaped from the 
acceleration zone and diffuse on scales ${\cal O}(100\,{\rm pc})$,
we do not address here the question of CR diffusion in the shock region. 
This allows us also to neglect the backreaction of CRs on the turbulent field, 
discussed e.g.\ in~\cite{Reville:2011qd}, which modifies the magnetic
field only in a thin layer in front of the shock.
We also only consider CRs with $E\gtrsim 1$\,TeV, which are those relevant for gamma ray observations above 100\,GeV.

\textit{CR diffusion and diffusion tensor}---Investigating the propagation of 
CRs close to their source requires to inject them localised in space, say at 
$\vec x=\vec 0$, and to propagate them in a given concrete realization of 
the turbulent field. We find that diffusion can be strongly anisotropic 
in a specific field realization, as long as the distance between CRs and the 
source is $\lsim l_{\max}$. This anisotropy is washed out averaging over many 
realizations of the turbulent field. Thus the correct procedure
to calculate the diffusion tensor in this case is to compute
$
D_{ij}^{(b)}= \frac{1}{2Nt}\sum_{a=1}^N x_i^{(a)}x_j^{(a)}
$
for $N$ particles (labeled by the subscript $a$) and injected at 
$\vec x=\vec 0$ in one single realization $b$. 
For each of the $M$ realizations one diagonalizes $D_{ij}^{(b)}$ and finds its
eigenvalues $d_{i}^{(b)}$. Then one averages the ordered eigenvalues, 
$d_1^{(b)}<d_2^{(b)}<d_3^{(b)}$, over the $M$ realizations,
$d_{i}= \frac{1}{M}\sum_{b=1}^M d_i^{(b)}$.


\begin{figure}
\includegraphics[width=\linewidth,angle=0]{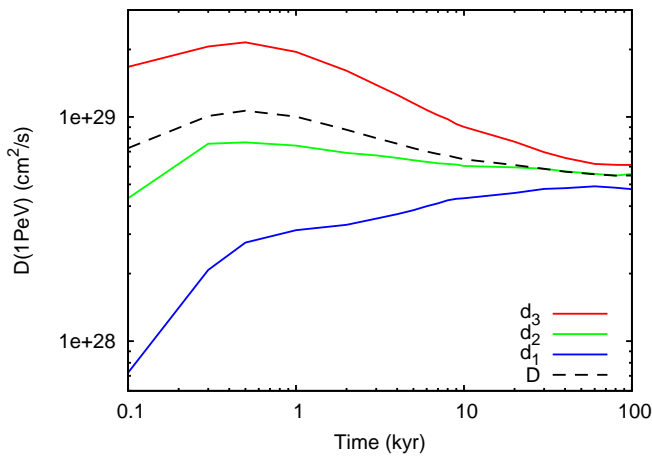}
\caption{
Eigenvalues $d_i$ (solid lines) of the diffusion tensor $D_{ij}=\langle x_ix_j\rangle/(2t)$ 
together with the average diffusion coefficient $D$ (dashed line)
as function of time $t$. For $B_{\rm rms}=4\,\mu$G, $l_{\max}=150$\,pc, $\alpha=5/3$
and CR energy $E=10^{15}$\,eV. 
\label{fig1}}
\end{figure}

In Fig.~\ref{fig1}, we show the three eigenvalues $d_i$
of $D_{ij}$ as function of time for the case of CRs with energy 
$E=10^{15}$\,eV. We used $M=10$ realizations, propagating for each 
$N=10^{4}$ particles. At early times $t \lesssim t_\ast$, the diffusion
tensor is strongly anisotropic, and the ratio $d_3/d_1$ between its largest and
smallest eigenvalue can reach a factor of a few hundreds in some turbulent 
field realizations, while it is on average around a factor of a few tens. 
For $t\gtrsim t_\ast$, CRs propagate more and more isotropically, 
approaching the predictions of the diffusion approximation for a purely 
isotropic turbulent field.

Modes with large scale variations contain most of the power in 
Kolmogorov turbulence, compared to the smaller scale variation modes 
which are responsible for the diffusion of TeV--PeV CRs.
Particles see modes with large variation scales as local 
uniform fields and diffuse therefore anisotropically.
Once a sufficient fraction of particles moved beyond $|{\vec x}|\sim l_{\max}$, 
the anisotropies of different ``cells'' of size $l_{\max}^3$ are averaged out
and the CR densities around sources tend towards the limit predicted by the diffusion 
approximation. Adding a large scale regular field $B_0$ on top 
of the turbulence would increase the anisotropies. In this case, CRs are
known to diffuse faster in the direction of $B_0$ than in the perpendicular 
direction.

The dashed line in Fig.~\ref{fig1} represents the average diffusion
coefficient $D= 1/M \sum_{b=1}^M D^{(b)}$, with $D^{(b)}$ defined as
$D^{(b)}= 1/(6Nt) \sum_{a=1}^N {\vec x}^{(a)}\cdot{\vec x}^{(a)}$
for the magnetic field realization $b$. For the largest times numerically
reachable, $t=10^5$\,yr, the eigenvalues
in Fig.~\ref{fig1} and the average $D$ approach a
common value, $D \ap 5.5 \times 10^{28}$\,cm$^2/$s. 
Interestingly, $2\times 10^4$\,yr corresponds to 
$\langle {\vec r}^2\rangle^{1/2}=\sqrt{6Dt}\ap l_{\max}$ valid in the
isotropic limit.
We verified that the limiting value for the average $D$
agrees with the value of the diffusion
coefficient computed using CRs with random starting positions.
It is consistent, among others, with the computations of~\cite{Casse:2001be} for pure random fields. 
The value of $D(E)$ in our Galaxy is currently only known within a factor 
$\ap 50$ at $E_0\sim 10$\,GeV, and $5.5 \times 10^{28}$\,cm$^2/$s is in the 
acceptable range extrapolating $D(E)$ to $E=10^{15}$\,eV for a Kolmogorov 
spectrum ~\cite{Maurin:2002ua,Timur:2011vv}. 
It is a factor $\ap 4$ smaller than the value 
used in Ref.~\cite{galprop}.

Earlier works~\cite{JG99,Casse:2001be,DeMarco:2007eh,Parizot:2004wh} did not 
report anisotropic diffusion for several reasons: Diffusion coefficients were computed averaging over
several configurations, with random initial positions for particles, or the considered
space or time scales were too large. For instance, \cite{JG99} and~\cite{Casse:2001be}
find isotropic diffusion in the limit of a vanishing uniform field, $B_0\ll B_{\rm rms}$. In both works 
the diffusion coefficient are calculated averaging over many realisations of the turbulent field. As a result, any anisotropy is averaged out, if the random field is isotropic. We also stress that 
$t_\ast$ is much larger than the transition time $\tau_{\rm diff}\sim 4D/c^2$
from the ballistic to the diffusive regime~\cite{Parizot:2004wh}.

\begin{figure*}[!t]
  \centerline{\includegraphics[width=0.333\textwidth]{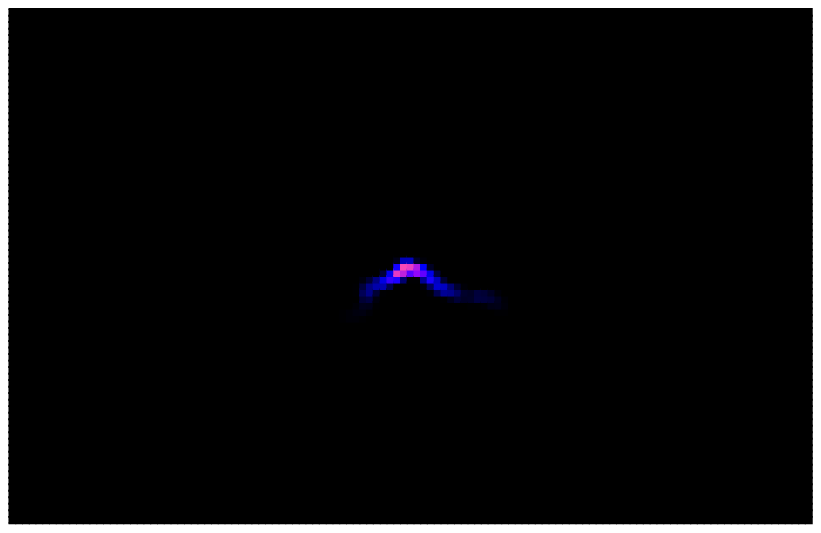}
              \hfil
              \includegraphics[width=0.333\textwidth]{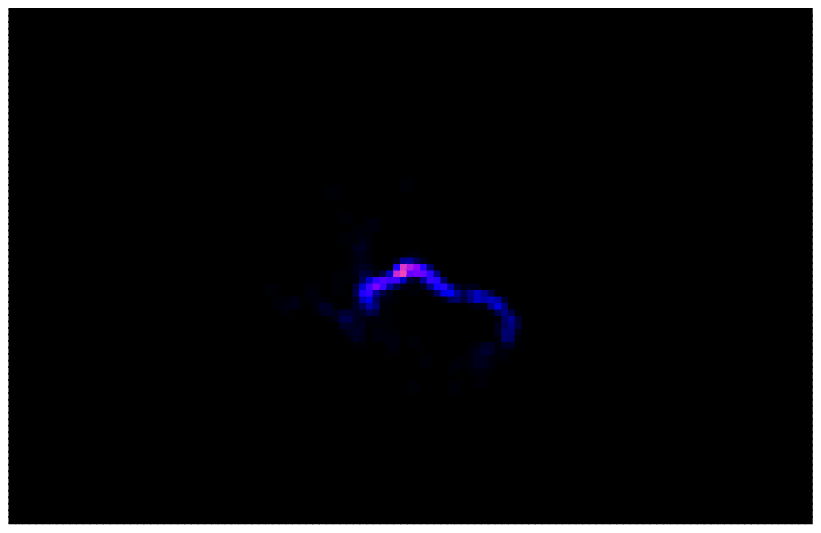}
              \hfil
              \includegraphics[width=0.333\textwidth]{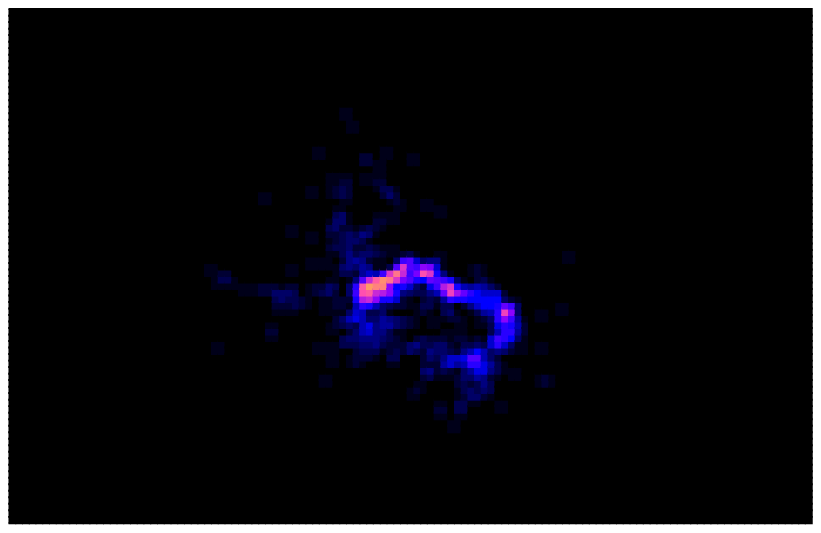}
             }
  \centerline{\includegraphics[width=0.333\textwidth]{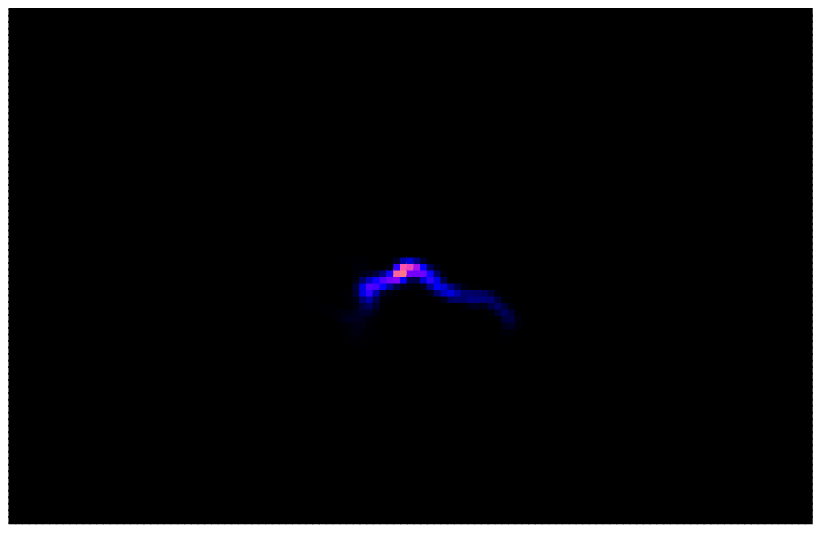}
              \hfil
              \includegraphics[width=0.333\textwidth]{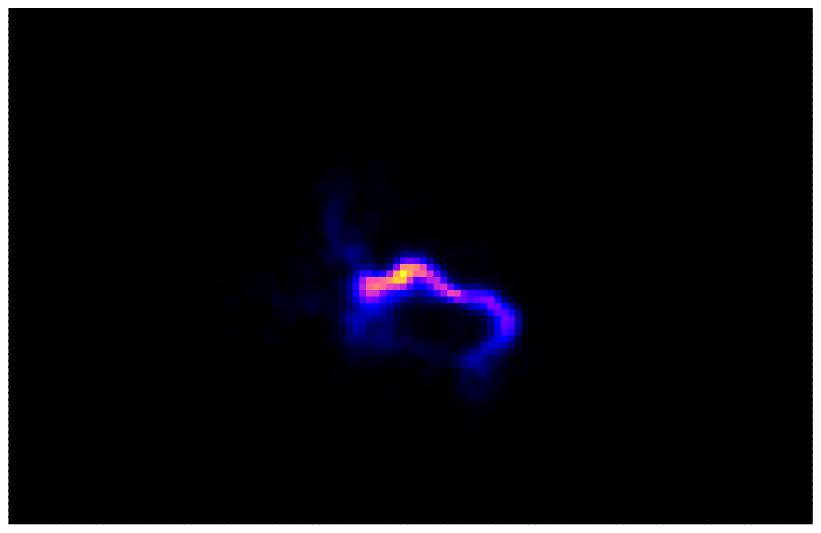}
              \hfil
              \includegraphics[width=0.333\textwidth]{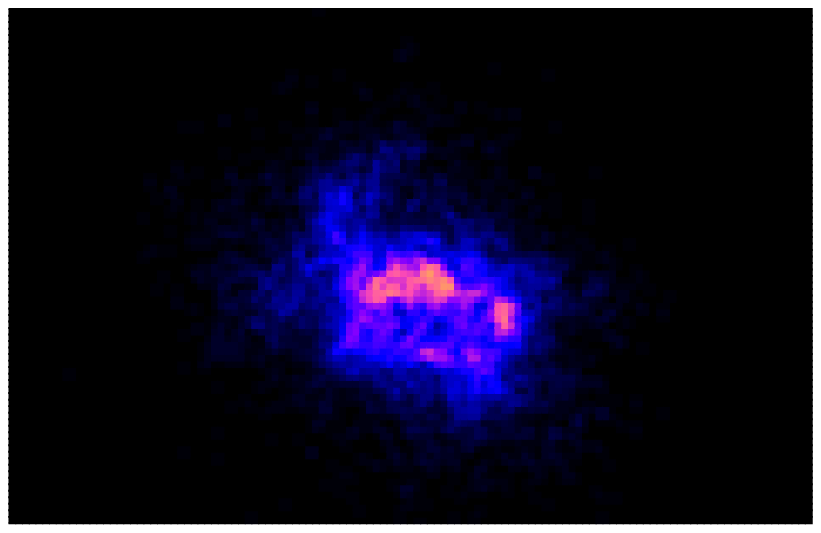}
             }
  \centerline{\includegraphics[width=0.333\textwidth]{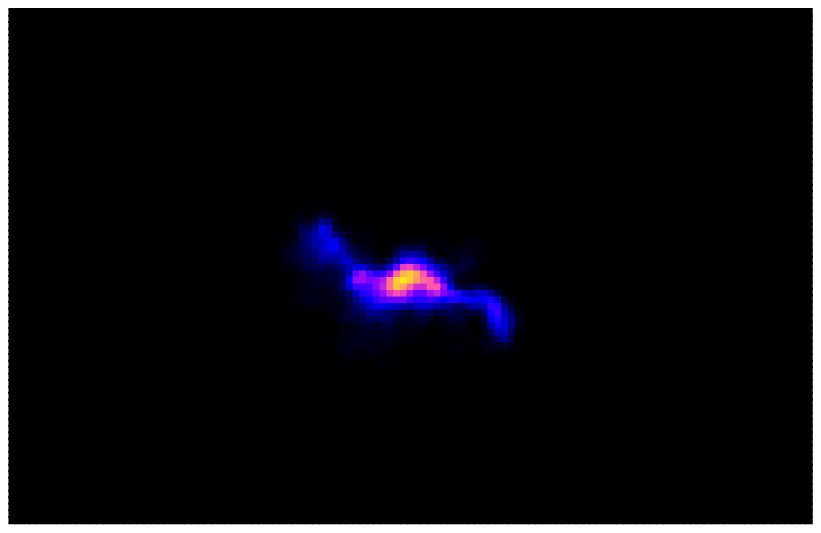}
              \hfil
              \includegraphics[width=0.333\textwidth]{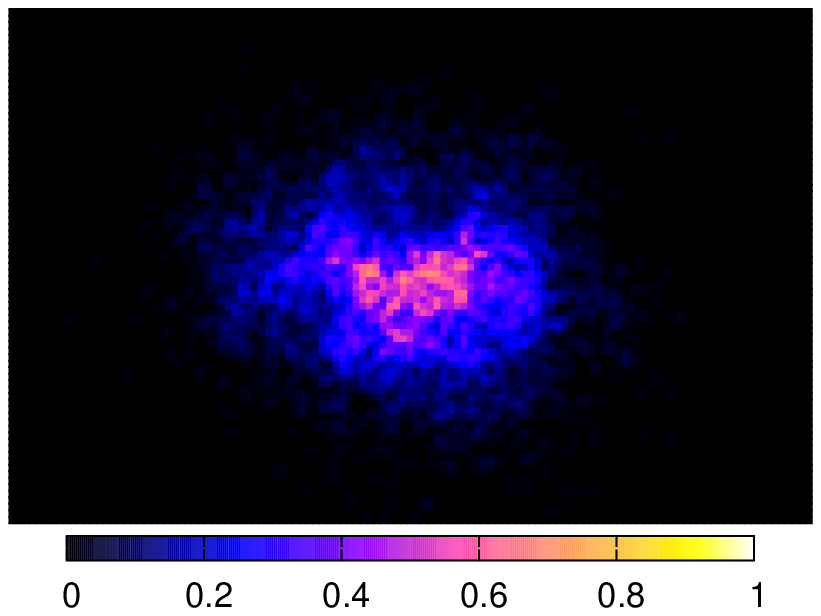}
              \hfil
              \includegraphics[width=0.333\textwidth]{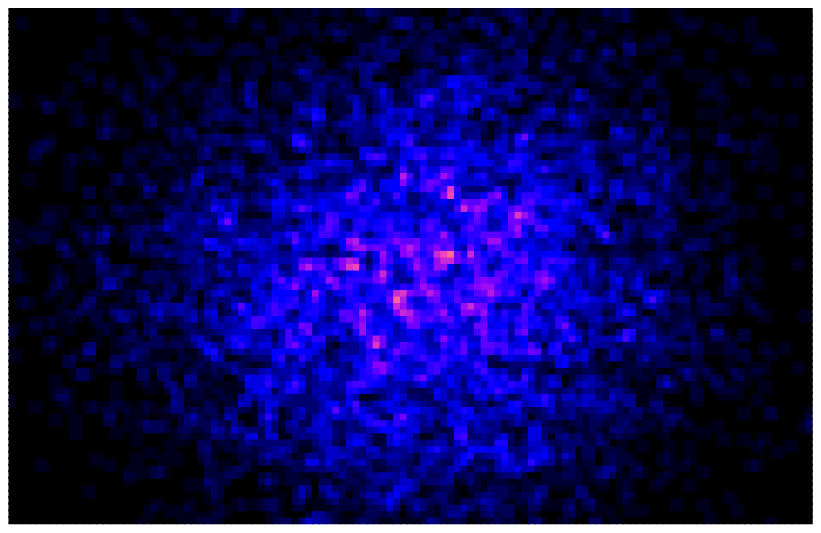}
             }
  \caption{Relative cosmic ray densities around their source projected in the panel planes, for energies $E=100$\,TeV \textit{(upper row)}, $1$\,PeV \textit{(middle row)}, $10$\,PeV \textit{(lower row)} and times $t=500$\,yr \textit{(left column)}, $2$\,kyr \textit{(middle column)}, $7$\,kyr \textit{(right column)}. Same field realization in each panel. Each panel corresponds to a $600\,{\rm pc}\times 400\,{\rm pc}$ field-of-view, with the source located in the center.}
  \label{fig2}
\end{figure*}


\textit{Cosmic ray intensity and extrapolation to low E.}---
In the middle row of Fig.~\ref{fig2}, we show the projection of the number density of 1\,PeV CRs on an arbitrarily chosen plane of size $600\,{\rm pc}\times 400\,{\rm pc}$ containing the injection 
point ${\vec x}={\vec 0}$ in the center. We consider here one given realization of the turbulent field, out of the ten used for Fig.~\ref{fig1}. The diffusion is confirmed to be strongly anisotropic at early times, 500\,yr (left) and 2000\,yr (middle panel), and even strongly filamentary. At 7000\,yr (right panel), the distribution of CRs slowly tends towards the spherical limit expected for true isotropic diffusion. Most of the nine other configurations display similar filamentary structures at early times. For a few of them, no thin filaments are visible, but the CR distribution around the source is still asymmetric, showing wind-like structures, also visible for some of the observed sources~\cite{HESS}.

\begin{figure*}[!t]
  \centerline{\includegraphics[width=0.33\textwidth]{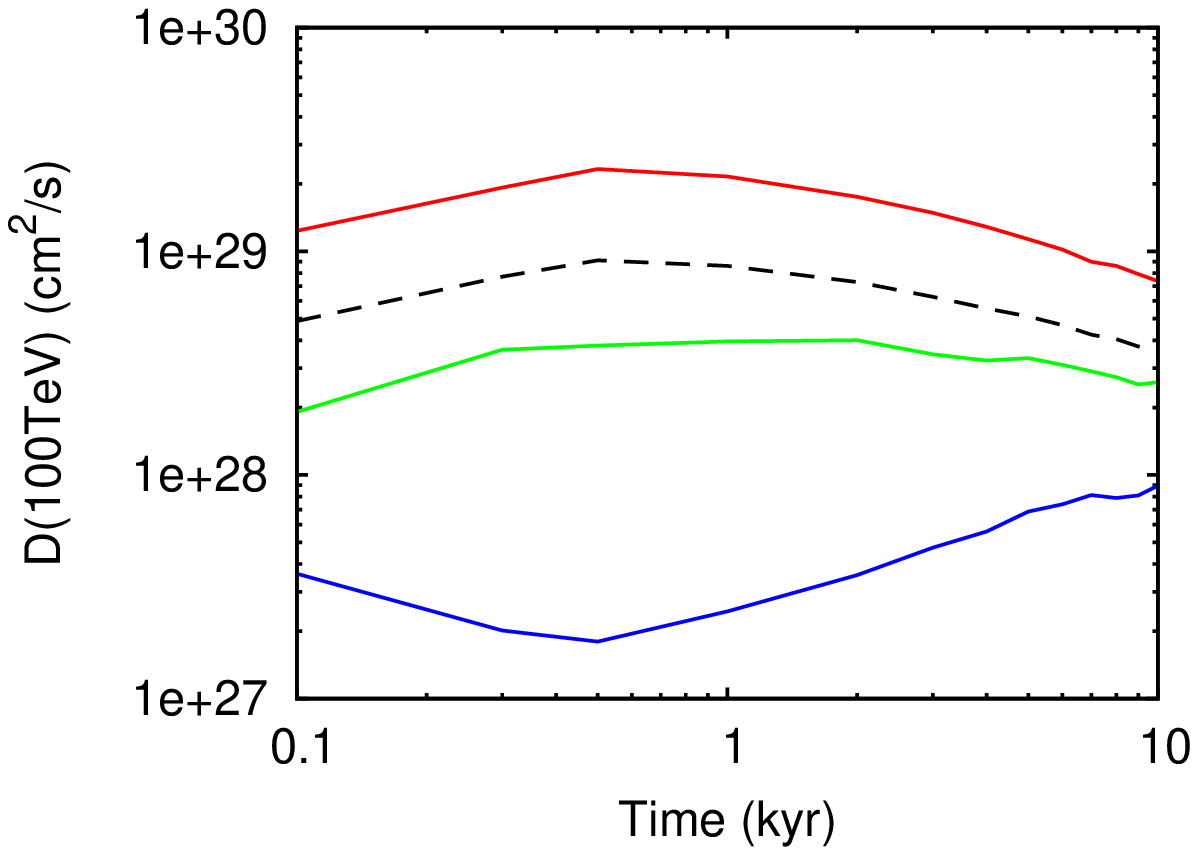}
              \hfil
              \includegraphics[width=0.33\linewidth]{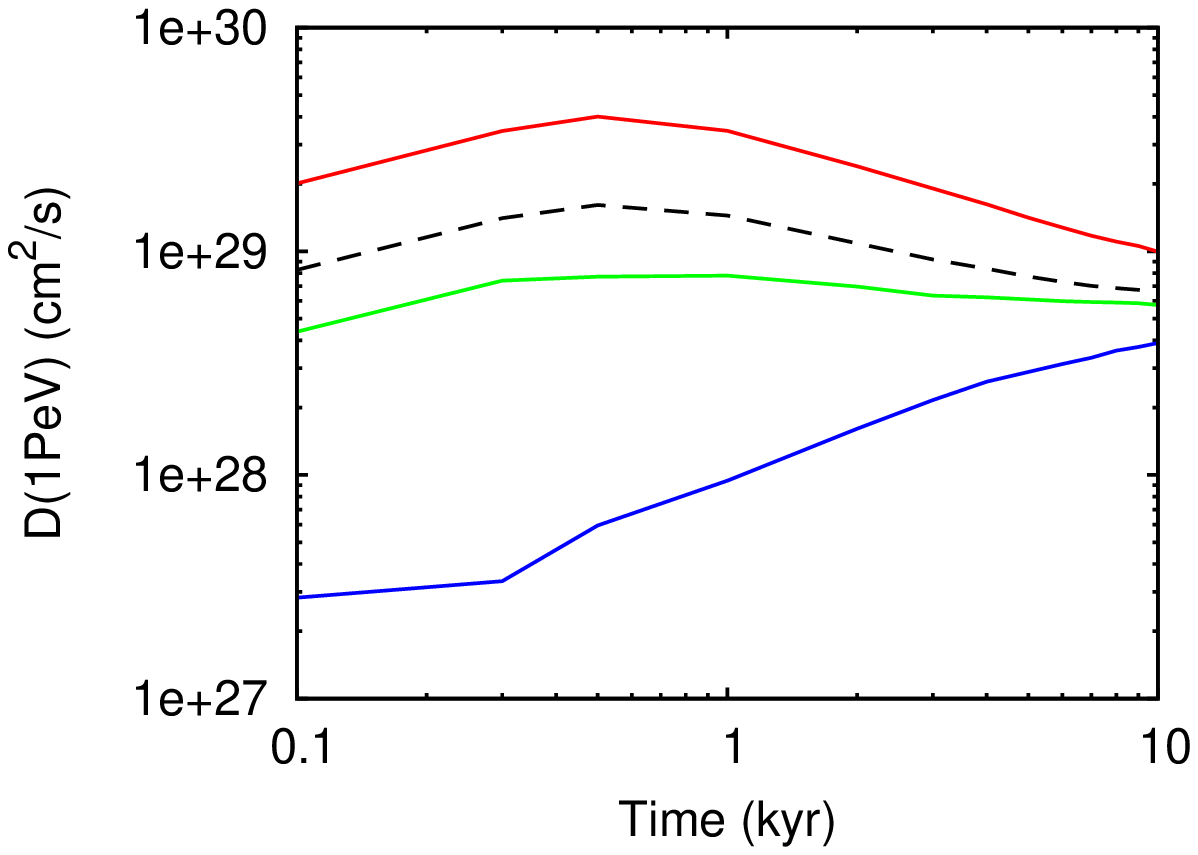}
              \hfil
              \includegraphics[width=0.33\textwidth]{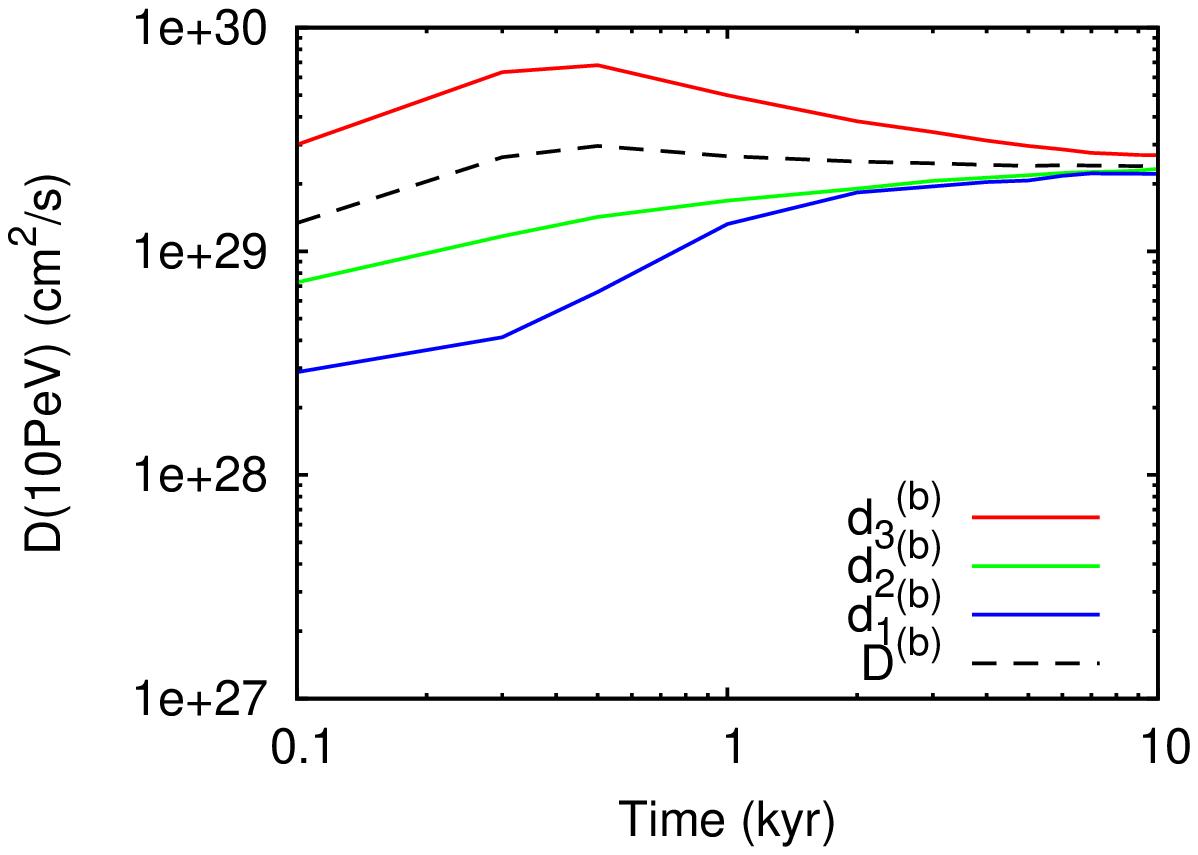}
             }
  \caption{Eigenvalues $d_i^{(b)}$ of the diffusion tensor $D_{ij}$ as function of time $t$ for energies $E=100$\,TeV \textit{(left)}, $1$\,PeV \textit{(middle)} and $10$\,PeV \textit{(right)}. Same field realization as in Fig.~\ref{fig2}. Dashed lines for the average diffusion coefficient $D^{(b)} \simeq D$.}
  \label{fig3}
\end{figure*}

The upper and lower rows of Fig.~\ref{fig2} present results for $E=100$\,TeV 
and $E=10$\,PeV, respectively. A comparison of the three rows shows that the period
of anisotropic diffusion lasts longer for CRs with lower energy.
For instance, the panel with $t=2$\,kyr and $E=1$\,PeV is very
similar to the one with $t=7$\,kyr and $E=100$\,TeV, which suggests that 
the expected scaling  $t\propto 1/D(E) \propto E^{-1/3}$ also holds in the case of anisotropic diffusion.
To determine this scaling law more quantitatively, we plot the values of $d_1^{(b)}$, $d_2^{(b)}$, $d_3^{(b)}$ and $D^{(b)}$ as function of time for given realizations $b$ in Fig.~\ref{fig3} and look for times with similar $d_3^{(b)}/d_1^{(b)}$:
Numerically we find $t_\ast \propto E^{-\gamma}$ with $\gamma=0.25$--$0.5$, 
i.e.\ a value of $\gamma$ consistent with the theoretical expectation.

\textit{Transition time.}---For the parameters of Fig.~\ref{fig1} one 
estimates $t_{\ast}\sim10^4$\,yr. This value is mostly determined by 
$l_{\max}$, and we find that the naive expectation $t_{\ast} \propto l_{\max}^2$ 
holds in a first approximation. Reducing $l_{\max}$ by a factor 6, to 25\,pc, 
the transition happens at 200--$300\simeq10^4/6^2$\,yr in all ten 
tested configurations. Therefore, our numerical results suggest that the 
diffusion approximation predictions become valid at
\be \label{tast}
t_{\ast} \sim 10^4 \,{\rm yr}\; \left( l_{\max}/150\,{\rm pc}\right)^{\beta}
\left( E/{\rm PeV} \right)^{-\gamma}
\left(  B_{\rm rms} /4\,{\rm \mu G} \right)^{\gamma}
\ee
with $\beta \simeq 2$ and $\gamma=0.25$--$0.5$ for Kolmogorov turbulence. $B_{\rm rms} \simeq 4\,\mu G$ corresponds to the estimated local value around the Earth from unpolarized synchrotron radiation data~\cite{Beck:2008ty}. 
However, the large uncertainties of the turbulent field parameters, in particular of $l_{\max}$, imply that $t_{\ast}$ may  differ significantly from $10^4$\,yr for PeV CRs.

The spectral index $\alpha$ does not have a strong impact on $t_{\ast}$. In contrast, it influences the ratio $d_3/d_1$ at early times. For a Bohm spectrum, we find that the anisotropy at $t \lesssim t_{\ast}$ is reduced. Indeed, more power is concentrated in small scale modes than in a Kolmogorov spectrum, resulting in a better isotropization of PeV CRs. After averaging over five realizations of the field, we find a factor $d_3/d_1 \ap 3$--$4$ at early times. Out of these five configurations, only one is Kolmogorov-like with $d_3^{(b)}/d_1^{(b)} \sim 10$ and visible filaments. The others are just slightly anisotropic.

\textit{Gamma-ray emission.}---High energy protons can
scatter on protons of the interstellar gas, producing secondaries
which in turn decay into photons. We simulate cross sections and the final 
state of proton-proton interactions using QGSJET-II~\citep{qgsjet}, while 
we use SIBYLL~2.1~\citep{sibyll1.7} for the subsequent decays of 
unstable particles. The emission of secondaries is strongly forward 
beamed and CRs in filaments have an anisotropic distribution of momenta. We 
may expect that the emitted $\gamma$-rays are a good tracker of 
the underlying CR anisotropies.

\begin{figure}
\includegraphics[width=0.49\linewidth,angle=0]{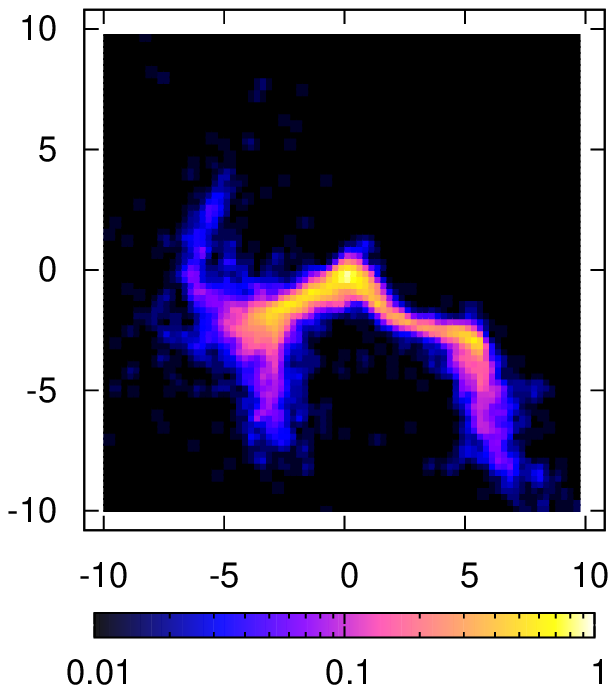}
\includegraphics[width=0.49\linewidth,angle=0]{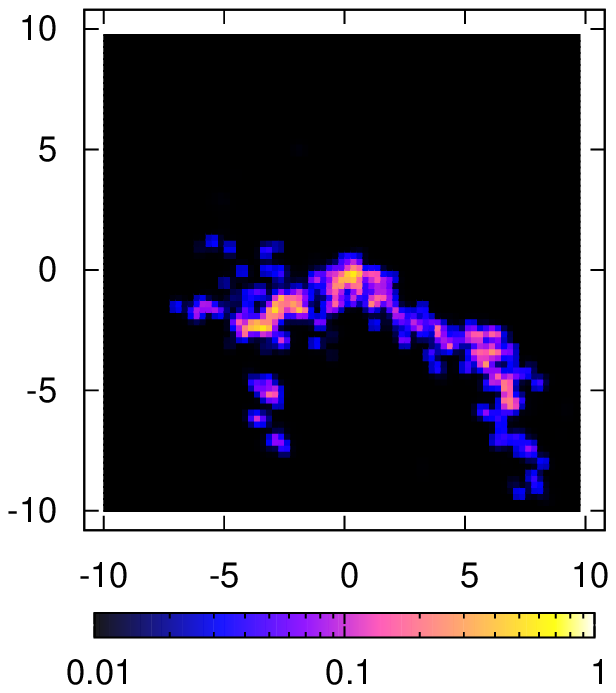}
\caption{\textit{Left panel:} Relative cosmic ray densities along the lines-of-sight as seen from a specific observer located 500\,pc away from the source. $E=1$\,PeV and $t=1$\,kyr, for a given magnetic field realization. Box size is $20^{\circ} \times 20^{\circ}$, with the source at (0,0). \textit{Right panel:} Corresponding relative surface brightness in $\gamma$-rays with energies $E_{\gamma} \geq 100$\,GeV.}
\label{fig4}
\end{figure}

For simplicity, we assume a uniform gas density
around the source. Then we place an observer at 500\,pc distance from the source
and integrate the photon flux emitted along the line-of-sight 
towards the observer. We model the observer as a sphere of 5\,pc,
which is the smallest size providing reasonable statistics and
introducing only a small amount of artificial 'fuzziness'. The
resulting source image is shown in the right panel of Fig.~\ref{fig4}. The comparison to the
corresponding CR intensity (left panel) shows that the
latter can be used to predict the shape of the gamma ray halo. Note that
gamma rays emitted via Compton scattering by electrons would
also display such anisotropic patterns.

\textit{Conclusions.}---\label{conclusions}
We studied the diffusion of TeV--PeV CRs on scales $l$ smaller or comparable to 
the largest scales of magnetic field fluctuations, $l\lsim l_{\max}$.
The propagation of such CRs close to their sources has to be studied in
single realizations of the turbulent field. We showed that CRs 
diffuse  anisotropically at early times $t\lesssim t_\ast$, with $t_\ast$ from Eq.~(\ref{tast}), leading to a filamentary structure of the CR density around their sources. Turbulent field modes with variation scales much larger than the Larmor radius of CRs are responsible for this anisotropic diffusion regime.
This effect can explain the observations of irregular gamma-ray halos~\cite{NS12,HESS,VERITAS} around CR proton and electron sources. If CRs propagate distances $l\gsim l_{\max}$, these anisotropies are averaged out. CR densities around sources become isotropic and tend towards those expected from the diffusion approximation.


\acknowledgments
GG acknowledges support from the Research Council of Norway through 
an Yggdrasil grant.

\end{document}